\documentclass[acmtog]{acmart}
\acmSubmissionID{1146}

\usepackage{booktabs} 

\usepackage[ruled]{algorithm2e} 

\PassOptionsToPackage{table,dvipsnames}{xcolor}
\usepackage{caption}
\usepackage{subcaption}
\usepackage{xcolor}
\usepackage{colortbl}
\usepackage{bm}
\usepackage{amsmath}
\usepackage{array}
\usepackage{multirow}

\DeclareMathOperator*{\argmin}{arg\,min}

\usepackage[capitalize]{cleveref}
\crefname{section}{Sec.}{Secs.}
\Crefname{section}{Section}{Sections}
\Crefname{table}{Table}{Tables}
\crefname{table}{Tab.}{Tabs.}
\definecolor{Cerulean}{rgb}{0.0, 0.48, 0.65}
\definecolor{myred}{rgb}{1, 0.6, 0.6}
\definecolor{myyellow}{rgb}{1,1, 0.6}
\definecolor{myorange}{rgb}{1, 0.8, 0.6}
\definecolor{mycolor_blue}{HTML}{E7EFFA}
\definecolor{mycolor_green}{HTML}{E6F8E0}
\definecolor{mycolor_gray}{HTML}{ECECEC}
\definecolor{pearDark}{HTML}{2980B9}
\newcommand{\tablefirst}[0]{\cellcolor{pearDark!20}}
\newcommand{\tablesecond}[0]{\cellcolor{mycolor_green}}

\SetAlFnt{\small}
\SetAlCapFnt{\small}
\SetAlCapNameFnt{\small}
\SetAlCapHSkip{0pt}

\copyrightyear{2025}
\acmYear{2025}
\setcopyright{cc}
\setcctype{by}
\acmConference[SA Conference Papers '25]{SIGGRAPH Asia 2025 Conference Papers}{December 15--18, 2025}{Hong Kong, Hong Kong}
\acmBooktitle{SIGGRAPH Asia 2025 Conference Papers (SA Conference Papers '25), December 15--18, 2025, Hong Kong, Hong Kong}\acmDOI{10.1145/3757377.3763835}
\acmISBN{979-8-4007-2137-3/2025/12}

\begin{document}
\title{Light-SQ: Structure-aware Shape Abstraction with Superquadrics for Generated Meshes}

\author{Yuhan Wang}
\affiliation{%
 \institution{S-Lab, Nanyang Technological University}
 \country{Singapore}}
\author{Weikai Chen} \authornote{Corresponding authors}
\author{Zeyu Hu}
\author{Runze Zhang}
\author{Yingda Yin}
\author{Ruoyu Wu}
\author{Keyang Luo}
\author{Shengju Qian}
\author{Yiyan Ma}
\author{Hongyi Li}
\author{Yuan Gao}
\author{Yuhuan Zhou}
\author{Hao Luo}
\author{Wan Wang}
\author{Xiaobin Shen}
\author{Zhaowei Li}
\author{Kuixin Zhu}
\author{Chuanlang Hong}
\author{Yueyue Wang}
\author{Lijie Feng}
\author{Xin Wang} \authornotemark[1]
\affiliation{%
 \institution{LIGHTSPEED}
 \country{China}
 }
\author{Chen Change Loy}
\affiliation{%
 \institution{S-Lab, Nanyang Technological University}
 \country{Singapore}}

\renewcommand\shortauthors{Wang et al.}

\begin{abstract}
In user-generated-content (UGC) applications, non-expert users often rely on image-to-3D generative models to create 3D assets. 
In this context, primitive-based shape abstraction offers a promising solution for UGC scenarios by compressing high-resolution meshes into compact, editable representations.
Towards this end, effective shape abstraction must therefore be structure-aware, characterized by low overlap between primitives, part-aware alignment, and primitive compactness.
We present Light-SQ, a novel superquadric-based optimization framework that explicitly emphasizes structure-awareness from three aspects. (a) We introduce SDF carving to iteratively udpate the target signed distance field, discouraging overlap between primitives. (b) We propose a block-regrow-fill strategy guided by structure-aware volumetric decomposition, enabling structural partitioning to drive primitive placement. (c) We implement adaptive residual pruning based on SDF update history to surpress over-segmentation and ensure compact results. 
In addition, Light-SQ supports multiscale fitting, enabling localized refinement to preserve fine geometric details. To evaluate our method, we introduce 3DGen-Prim, a benchmark extending 3DGen-Bench with new metrics for both reconstruction quality and primitive-level editability.
Extensive experiments demonstrate that Light-SQ enables efficient, high-fidelity, and editable shape abstraction with superquadrics for complex generated geometry, advancing the feasibility of 3D UGC creation.
\emph{Project Page: \href{https://johann.wang/Light-SQ/}{https://johann.wang/Light-SQ/}}.
\end{abstract}

%
%


%
%


\begin{teaserfigure}
    \centering
    \vspace{-2pt}
    \includegraphics[width=0.98\textwidth]{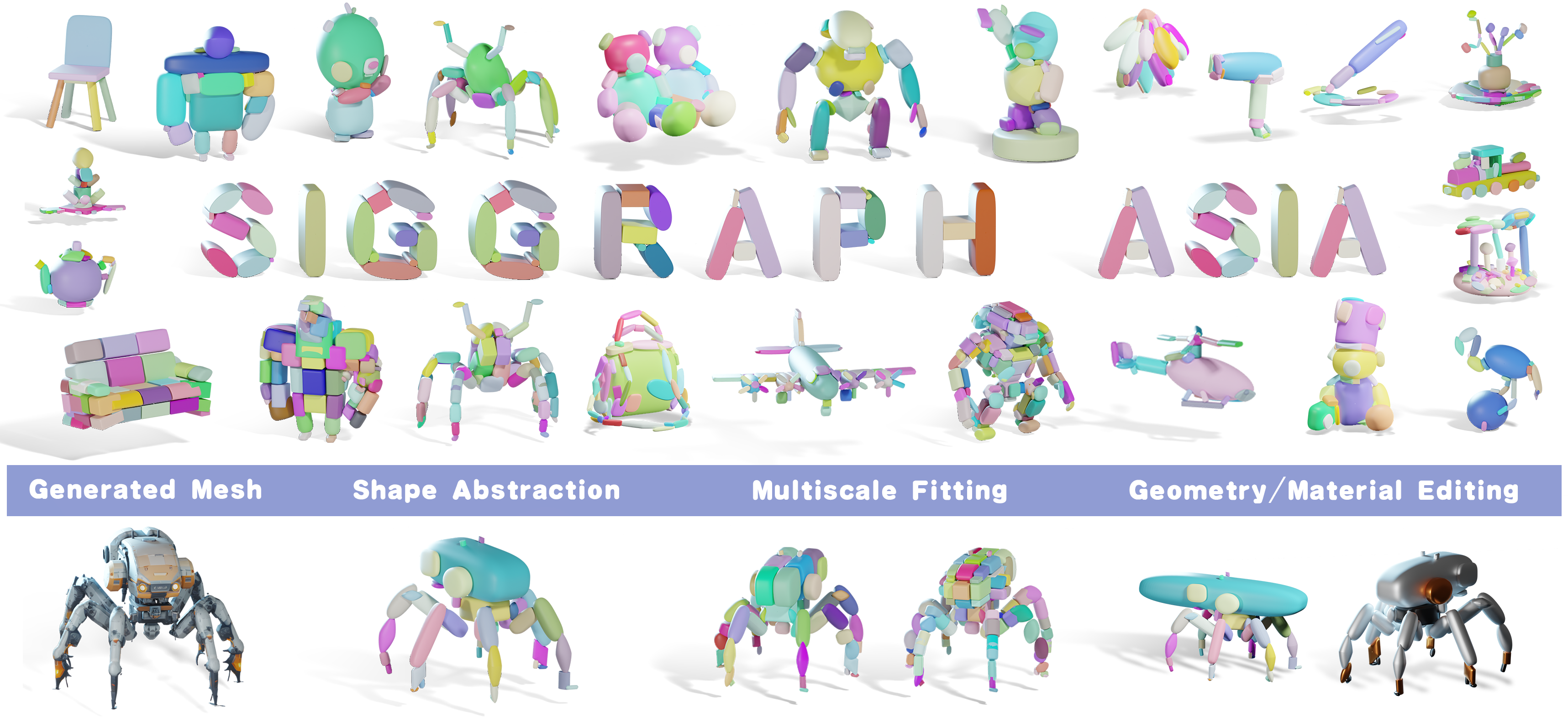}
    \vspace{-6pt}
    \caption{We introduce \textbf{Light-SQ}, a superquadrics-based shape abstraction method tailored for generated 3D meshes and well-suited for UGC scenarios.}
    \label{fig:teaser}
\end{teaserfigure}

\maketitle

\section{Introduction}
\label{sec:intro}

Customizable 3D asset creation is a fundamental requirement in user-generated-content (UGC) platforms, yet remains largely inaccessible to non-expert creators.
Recent advances in image-to-3D generation \cite{zhang2024clay, li2025triposg, yang2024hunyuan3d1, zhao2025hunyuan3d2, lai2025unleashing, chen2025mar3d} have significantly lowered the entry barrier, allowing users to generate high-quality 3D meshes from simple image prompts.
However, the resulting meshes are typically over-tessellated, structurally unorganized, and difficult to edit, posing critical challenges in downstream applications such as animation, rigging, and interactive content creation.

To address these limitations, primitive-based shape abstraction has emerged as a promising solution. By converting bulky triangle meshes into a compact set of analytic primitives, this approach offers two key benefits: 1) substantial storage reduction, from megabytes to kilobytes; and 2) improved editability, as each primitive serves as an intuitive manipulation handle.  
Our objective is to extend these benefits to generated 3D assets, transforming coarse, unstructured meshes into compact, editable primitive representations, for a robust image-to-primitives pipeline.

\begin{figure}[t]
    \centering
    \includegraphics[width=\columnwidth]{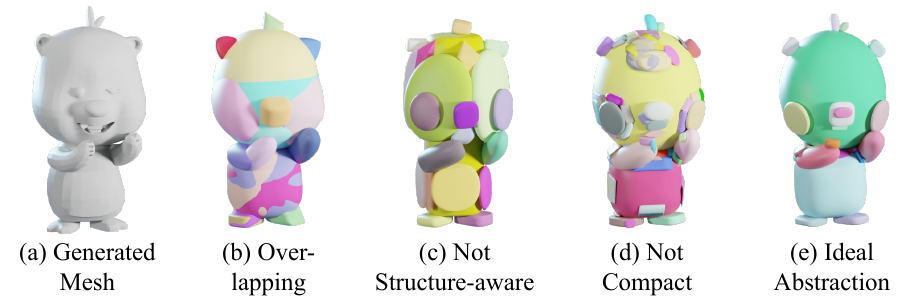}
    \caption{Given the input mesh, (b), (c), and (d) exhibit, respectively, excessive primitive overlap, structural inconsistency, and excessive fragmentation. In (c), the centric primitive spans the head and the body.}
    \vspace{-0.2cm}
    \label{fig:ugc_standards}
\end{figure}

For UGC scenarios, an ideal primitive abstraction should satisfy two high-level goals: \emph{fidelity} and \emph{editability}.
While ``fidelity" refers to the preservation of salient shape structures, ``editability" relies on satisfying three structure-aware criteria: (a) \emph{Low overlap}. Each primitive should occupy a distinct, low-overlapping spatial region; (b) \emph{Structure awareness}. Primitives should conform to coherent volumetric partitions, avoiding cross-structure primitive placement; (c) \emph{Compactness}. Excessive segments would hinder downstream tasks such as collision detection, rigging, and editing. \cref{fig:ugc_standards} shows results that do not meet these criteria, along with an ideal abstraction.

Despite extensive research, existing abstraction techniques struggle with the irregularities and noise characteristic of generative geometry.
Learning-based approaches~\cite{smirnov2020deepshapepred, yang2021unsupervised,  paschalidou2020learning, li2024pasta, ye2025primitiveanything} directly regress primitive parameters from the inputs, but typically generalize poorly outside the training domain, which is often limited to clean, curated datasets such as ShapeNet.
Rule-based pipelines~\cite{lin2020seg, li2024lmp, li2025aissr} first segment the shape and then fit primitives to each segment. 
However, segmentation derived from generative meshes tend to be noisy -- either due to being trained on artist-annotated assets with clean boundaries or adapted from 2D image priors lacking 3D awareness. Furthermore, even semantically meaningful segments may exhibit geometric irregularities that prevent accurate approximation by a single primitive, ultimately resulting in inconsistent or fragmented decomposition.
Optimization-based techniques~\cite{wu2022primitivebayesian, liu2022emsrobust, liu2023marching, monnier2023differentiable, fedele2025superdec} avoid these pitfalls by iteratively fitting superquadrics to the geometry. While expressive, they often produce highly overlapped primitives, which compromises editability by eliminating clear ownership over spatial regions.

To address these issues, we present Light-SQ, a structure-aware superquadrics decomposition framework tailored for generated geometry in UGC settings. 
We leverage superquadric for their compact representation and closed-form expressiveness over novel objects. 
Inspired by Marching Primitives~\cite{liu2023marching}, we adopt truncated signed distance fields (TSDFs) as the fitting target to exploit rich geometric information.
To ensure structure awareness, we introduce three key components. 
First, we propose SDF carving, a volumetric exclusion mechanism that imposes explicit penalty over overlapping primitives to encourage spatial separation.
Second, we present a structure-aware alignment framework that leverages geometric-feature-driven volumetric partitioning and a block–regrow–fill strategy to guide superquadric fitting to conform to structural boundaries.
Third, to improve compactness, we track the SDF update history to classify residual primitives based on their geometric significance. Primitives below category-specific thresholds are discarded, preserving salient details while minimizing redundancy.

In addition, Light-SQ supports \emph{multiscale fitting}, allowing recursive refinement of coarse primitives to capture fine-grained geometry. This allows flexible balancing between abstraction detail and reconstruction quality, which prior methods lack.
Apart from the Light-SQ framework, we present 3DGen-Prim, a benchmarking suite for evaluating primitive decomposition on generated 3D geometry. Unlike existing protocols based on ShapeNet~\cite{Chang2015shapenet}, which features clean, curated meshes, 3DGen-Prim extends 3DGen-Bench~\cite{zhang20253dgenbench} with outputs from recent image-to-3D methods~\cite{zhao2025hunyuan3d2, li2025triposg}, and introduces metrics that assess both reconstruction fidelity and structural editability.

To summarize, our main contributions are as follows:
\begin{itemize}
\item We present Light-SQ, a superquadric decomposition algorithm that achieves high fidelity and structural awareness on generated geometry for the UGC scenarios. 
\item We introduce key algorithmic components, including SDF carving, structure-aware alignment, and adaptive residual pruning, which collectively promote structure-aware alignment and representation compactness.
\item Extensive experiments on 3DGen-Prim dataset demonstrate our significant advantages in fitting fidelity and editability.
\end{itemize}
\section{Related Work}
\label{sec:related}

\noindent\textbf{Optimization-based Shape Abstraction.}
Optimization-based approaches reconstruct 3D shapes by optimizing primitive parameters, typically superquadrics. 
State-of-the-art efforts~\cite{wu2022primitivebayesian, liu2022emsrobust} began by fitting primitives to point clouds. They treat the point cloud as an observation sampled from a probabilistic model upon superquadrics, and apply Maximum Likelihood Estimation (MLE) \cite{liu2022emsrobust} and Non-parametric Bayesian Inference \cite{wu2022primitivebayesian} respectively to solve their parameters.
Subsequent works explored fitting on other modalities. Marching-Primitives~\cite{liu2023marching} adopts the truncated signed distance field (TSDF) to avoid geometric ambiguities \cite{yang2021unsupervised}. \citet{monnier2023differentiable} incorporates superquadrics into a NeRF-like \cite{mildenhall2020nerf} reconstruction pipeline to fit multiview images.

Although existing methods have made significant progress in the accuracy of fitting, they lacked efforts to meet ``structure-aware requirements", as also noted in the concurrent work \cite{ye2025primitiveanything}. Light-SQ is the first superquadrics optimization algorithm that formally defines and explicitly emphasizes structural awareness.

\noindent\textbf{Learning-based Shape Abstraction.}
Learning-based approaches aim to employ a single neural network to predict the corresponding primitive representation from 2D or 3D inputs. 
\citet{tulsiani2017learningbyassembling} pioneered using a convolutional neural network for cuboid representation prediction with volume input. 
3D-PRNN~\cite{zou20173dprnn} takes a depth map as input and uses recurrent neural networks.
\citet{paschalidou2019superquadricsrevisted} extends the primitive type to superquadrics, and introduces point cloud reconstruction as the supervision. 
Subsequent works investigated the abstraction of other inputs such as distance fields \cite{smirnov2020deepshapepred} and the single-view image \cite{niu2018im2struct, paschalidou2020learning}, and explored the integration of frontier techniques, including unsupervised point cloud segmentation \cite{yang2021unsupervised} and autoregressive transformers \cite{li2024pasta}. 
Despite promising results in their data, the training source of these methods is typically confined to ShapeNet~\cite{Chang2015shapenet} (in some cases, only to a single subclass), resulting in poor generalization of the generated geometry.

Recently, SuperDec~\cite{fedele2025superdec} improved feed-forward predictions with an optimization refinement step. In parallel, PrimitiveAnything~\cite{ye2025primitiveanything} constructed an unprecedented 120K-scale shape abstraction dataset and trained an autoregressive network on point cloud inputs, achieving remarkable generalization on generated geometry. However, it falls short of the UGC standards outlined in \cref{sec:intro} and does not support multiscale decomposition.

\noindent\textbf{Rule-based Shape Abstraction.}
Rule-based methods typically bind each primitive with a segmentation part. SEG-MAT~\cite{lin2020seg} uses the medial axis transform to partition a mesh into skeletal regions and represent each with a cuboid. LMP~\cite{li2024lmp} introduces Shared Latent Membership, where deformable superquadrics serve as both segmentation priors and abstraction targets. More recently, AISSR~\cite{li2025aissr} aligns instance and semantic sparse representations to derive repeatable primitive templates. However, when applied to generated geometry, these segmentation schemes often fail to achieve high‐fidelity shape fitting.

\noindent\textbf{Approximate Convex Decomposition.}
Approximate convex decomposition (ACD) splits 3D geometry into a minimal set of pseudo-convex parts. CoACD~\cite{wei2022coacd_approximate} introduces a split-and-merge decomposition scheme. \citet{andrews2024navigation} proposed navigation‐driven ACD, which allows more flexible decomposition requirements. In this paper, we developed an adaptive ACD algorithm tailored for structure-aware shape abstraction.
\section{Method}
\label{sec:method}

\begin{figure*}[t]
    \centering
    \includegraphics[width=0.95\textwidth]{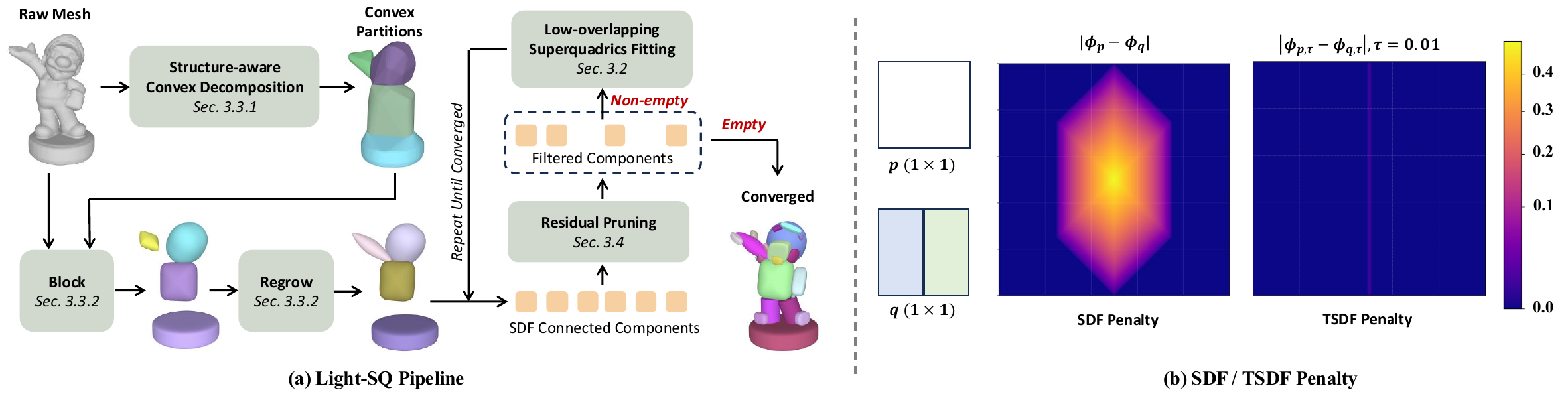}
    \vspace{-0.2cm}
    \caption{\textbf{Overview.} \emph{(a) Light-SQ pipeline.} Our method first employs block and regrow operations to satisfy the structural guidance of convex partitions, and then iteratively fills the SDF connected components until convergence, yielding a high-fidelity shape abstraction. \emph{(b) Visualization of SDF and TSDF difference.} If a closed shape ($p$) is abstracted in the form of two primitives ($q$), around the boundary between them, the SDF values deviate significantly from the ground truth, incurring an over-penalty that hinders the necessary decomposition. This issue is mitigated in the TSDF field.}
    \vspace{-0.4cm}
    \label{fig:preliminary}
\end{figure*}

In this section, we first introduce how to fit a signed distance field with superquadrics (\cref{subsec:preliminary}). Next, we describe how Light-SQ achieves structure-aware superquadric fitting from three perspectives: low-overlap (\cref{subsec:non_overlapping}), structure-aware alignment (\cref{subsec:semantic_alignment}), and compactness (\cref{subsec:pruning}). \cref{fig:preliminary}-(a) shows our fitting pipeline. Finally, we demonstrate the multiscale fitting capability enabled by our structure-aware framework (\cref{subsec:multiscale}).

\subsection{Preliminaries: Superquadrics Fitting on TSDF}
\label{subsec:preliminary}

A signed distance function (SDF) returns the distance from a given point to the closest surface, which is negative when it is inside:
\begin{equation}
    \phi(\bm{x})=SDF(\bm{x})=s:x\in\mathbb{R}^3,s\in\mathbb{R}
    \label{eq:sdf}
\end{equation}
Existing optimization-based approaches \cite{liu2023marching} use the truncated SDF (TSDF) $\phi_{\tau}(\bm{x})$ as the fitting target:
\begin{equation}
    \phi_{\tau}(\bm{x}) = TSDF(\bm{x}; \tau) = \mathrm{clamp}(SDF(\bm{x}), -\tau, +\tau)
    \label{eq:tsdf}
\end{equation}
The truncation parameter $\tau$ is typically set to the voxel edge length. Consequently, this field behaves almost like an \textbf{occupancy grid}, only providing local curvature information in regions adjacent to the surface. We find this setting reasonable, as it prevents overpenalty in the boundary region. We visualize it in \cref{fig:preliminary}-(b).

An axis-aligned superquadric can be represented by shape parameters $\epsilon_1,\epsilon_2\in[0,2]$ and scale parameters $a_x,a_y,a_z\in\mathbb{R}^{+}$ as
\begin{equation}
    f(\bm{x})=\left( \left( \frac{x}{a_x} \right)^{\frac{2}{\epsilon_2}} + \left( \frac{y}{a_y} \right)^{\frac{2}{\epsilon_2}} \right)^{\frac{\epsilon_2}{\epsilon_1}} + \left( \frac{z}{a_z} \right)^{\frac{2}{\epsilon_1}}=1,
    \label{eq:superquadrics}
\end{equation} where an arbitrarily positioned superquadric requires another three euler angles $\bm{e}\in\mathbb{R}^3$ and a translation vector $\bm{t}\in\mathbb{R}^3$ to describe the Euclidean transformation $g$. Given a superquadric parameterized by $\bm{\theta}=[\epsilon_1,\epsilon_2,a_x,a_y,a_z,g(\bm{e},\bm{t})]\in\mathbb{R}^{11}$ and a 3D point $\bm{x}$, we can use the signed radial distance function to approximate its SDF:
\begin{equation}
    \phi_{\bm{\theta}}(\bm{x})=\left( 1 - f^{-\frac{\epsilon_1}{2}}(g^{-1}\circ\bm{x}) \right)\|g^{-1}\circ\bm{x}\|_2
    \label{eq:sq_srdf}
\end{equation}

Superquadric fitting is performed on each SDF connected component. After a new superquadric is fitted, the connected component is usually subdivided into several smaller segments. To fit this new superquadric, existing methods start from an initial superquadric $\bm{\theta}$, and, at each iteration, optimize its parameters by fitting to the TSDF field within its local neighborhood $\bm{V}_a(\bm{\theta})$, then repeat this process with the updated superquadric parameter:
\begin{equation}
    \hat{\bm{\theta}}=\argmin_{\hat{\bm{\theta}}} \sum_{\bm{x}\in\bm{V}_a(\bm{\theta})} \lambda_{\bm{x}}\|\phi_{\bm{\hat{\theta}},\tau}(\bm{x})-\phi_{\tau}(\bm{x})\|_2^2
\end{equation}
Here $\lambda_{\bm{x}}$ controls the penalty weight applied to each voxel.

\subsection{Low-overlapping Superquadrics Fitting}
\label{subsec:non_overlapping}

In previous works \cite{liu2023marching}, after a superquadric is fitted, the voxels inside it will no longer be penalized during the subsequent fitting. This allows superquadrics to overlap with each other, providing more flexibility for filling the remained regions. However, when applied to generated geometry, the complex curvature distribution causes this strategy to produce a large number of highly overlapping superquadrics, which not only violates structural compactness but also hinders downstream editing and interaction in UGC scenarios.

We revisited the optimization process of a single superquadric. Previous works derive the penalty weights through maximum likelihood estimation. We find that the computational framework of $\lambda_{\bm{x}}$ can be relaxed to the following form:
\begin{equation}
    \lambda_{\bm{x}}=\frac{P(\phi_{\tau}(\bm{x}) | \phi_{\bm{\theta}, \tau}(\bm{x}))}{\underbrace{(\phi_{\tau}(\bm{x})<0)\cdot C\cdot(1-w)/w}_{\text{inside-voxel decay term}} + \underbrace{P(\phi_{\tau}(\bm{x}) | \phi_{\bm{\theta}, \tau}(\bm{x}))}_{\text{TSDF matching term}}}
    \label{eq:lambda}
\end{equation}
Here $w$ denotes the prior probability of any voxel covered by the current superquadric, which can be approximated as $1/\tilde{N}$ where $\tilde{N}$ is the expected number of superquadrics. $C$ is a weighting constant, and $P(\phi_{\tau}(\bm{x}) | \phi_{\bm{\theta}, \tau}(\bm{x}))$ follows a normal distribution. The TSDF matching term ranges between 0 and 1, while the inside‐voxel decay term is much larger. The motivation of this framework is to assign a much higher penalty to exterior voxels than to interior ones, thereby preventing a superquadric from expanding beyond the surface while still accommodating partial fitting within the interior regions.

Based on this property, we found that we can fulfill the low overlap requirement by iteratively updating the target SDF field. We convert the interior region of each fitted superquadric into the target shape’s exterior, thereby preventing subsequent superquadrics from encroaching on previously fitted regions. We call this operation \emph{SDF carving}, denoted by $\phi\setminus\phi_{\bm{\theta}}$. Specifically, $\phi\setminus\phi_{\bm{\theta}}$ updates the original SDF field of the target shape as:
\begin{equation}
 \phi(\bm{x})=\begin{cases}
-\phi_{\bm{\theta}}(\bm{x}), & \phi(\bm{x})<0 \wedge \phi_{\bm{\theta}}(\bm{x})<0,\\
\max(-\phi_{\bm{\theta}}(\bm{x}), \phi(\bm{x})), & \phi(\bm{x})<0 \wedge \phi_{\bm{\theta}}(\bm{x})>0,\\
\phi(\bm{x}), & \phi(\bm{x})>0
\end{cases}
\label{eq:sdf_carving}
\end{equation}

SDF carving does not alter the SDF values of voxels outside the surface. The SDF field within the superquadric can be directly replaced, while those still inside the surface after carving requires a comparison and update. After this, we re-clamp $\phi(\bm{x})$ as \cref{eq:tsdf} to obtain the updated TSDF field $\phi_{\tau}(\bm{x})$.

\subsection{Structure-Aware Alignment}
\label{subsec:semantic_alignment}

Compared to low overlap, achieving structure–aware alignment is substantially more challenging due to two primary issues. First, ``structure awareness'' lacks a precise and consistent definition. A natural approach is to define semantics through segmentation; however, most existing methods only produce surface-level labels from rendered images~\cite{tang2024samesh, yang2024sampart3d, liu2025partfield}, which do not translate meaningfully to voxels within the volume. In contrast, direct volumetric segmentation approaches~\cite{liu2025partfield, yang2025holopart} are much less common and often suffer from unstable results.
Second, even with a plausible structural segmentation, how to effectively leverage it for shape abstraction remains unclear.
Semantic regions identified by 3D segmentation models may not align well with the geometric assumptions of primitive fitting -- it is not guaranteed that each semantic part can be faithfully approximated by a single superquadric. This fundamental misalignment is also a key limitation for many rule-based abstraction pipelines.

Instead of relying on semantic learning, we adopt geometrical analysis to provide structural guidance. Our approach first introduces a structure-aware volumetric decomposition, followed by a three-phase convex segment-guided fitting strategy.

\subsubsection{Structure-Aware Convex Decomposition}
\label{subsubsec:decomposition}

To enable structure-aware volumetric decomposition, we build upon CoACD~\cite{wei2022coacd_approximate}, which formulates decomposition as a sequential decision process via MCTS. While effective, CoACD lacks structural guidance in plane sampling and relies on heuristic merging, often resulting in misaligned cuts and inconsistent part grouping. We address these issues by introducing structure-aware splitting and adaptive merging strategies to enhance structure alignment and geometric fidelity.

\begin{figure}[H]
    \centering
    \vspace{-0.4cm}
    \includegraphics[width=0.95\columnwidth]{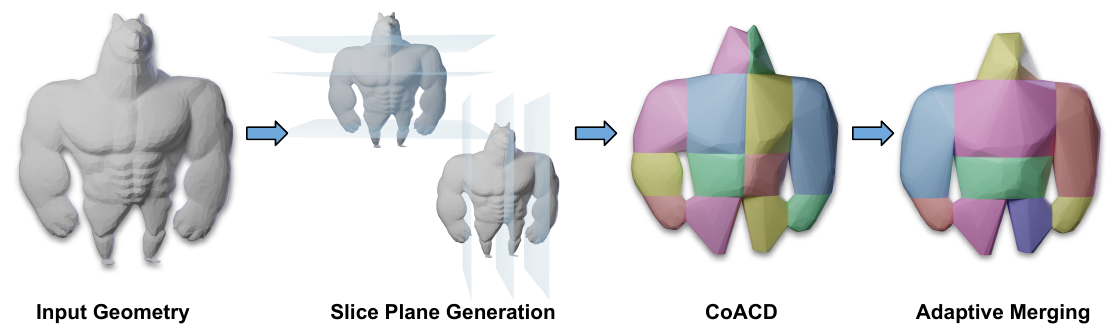}
    \vspace{-0.4cm}
    \caption{Structure-aware convex decomposition pipeline.}
    \vspace{-0.4cm}
    \label{fig:structure_aware}
\end{figure}

\noindent\textbf{Structure-Aware Slice Plane Generation.}
Instead of randomly sampling splitting planes, we precompute a fixed set of geometrically meaningful candidate planes using SDF-based volumetric analysis. Our method analyzes cross-sectional area variation and surface connectivity to select axis-aligned planes that are more likely to align with intrinsic geometric structures.
In particular, we evaluate each axis-aligned slice $i$ based on two geometric cues: (1) the second-order difference of the cross-sectional area $A_i$, and (2) the variation in the number of connected components $N_i$ on the slice.

The per-slice area $A_i = \sum_{j,k} \mathbf{1}\left[\text{SDF}(i,j,k) < 0\right],$ where $\mathbf{1\left[ \cdot \right]}$ returns 1 if the condition holds, and 0 otherwise. 
We define its second-order difference (window size 3) to capture abrupt area transitions:
\begin{equation}
M_i = \sum_{j=1}^{3} \left( A_{i-j} + A_{i+j} \right) - 2 \left( A_{i-1} + A_i + A_{i+1} \right)
\end{equation}

Let $N_i$ denote the number of connected components formed by interior voxels on slice $i$. We compute the discrete component variation
$\Delta N_i = |N_i - N_{i-1}|.$
The final structural saliency score is:
\begin{equation}
S_i = \alpha \cdot \tilde{M}_i + (1 - \alpha) \cdot \tilde{\Delta N}_i,
\end{equation}

\noindent where $\tilde{M}_i$ and $\tilde{\Delta N}_i$ are normalized scores, and $\alpha \in [0,1]$ balances area change and connectivity jumps. We then select the top-$K$ scoring slice indices as candidate planes, with a minimum spacing constraint $\delta$ to ensure diversity.

\noindent\textbf{Adaptive Merging based on Geometric Continuity.} To ensure meaningful part decomposition and avoid over-segmentation and erroneous merging, we propose an adaptive merging strategy based on geometric continuity. This strategy jointly considers curvature similarity and volumetric alignment between adjacent convex parts.

To compute curvature continuity, we denote $\Gamma$ as the shared interface between two convex parts $C_1$ and $C_2$. For each point $p \in \Gamma$, let $H_1(p)$ and $H_2(p)$ represent the mean curvature evaluated on the respective sides. The curvature continuity score is defined as:
\begin{equation}
S_{\text{curv}}(C_1, C_2) = 1 - \frac{1}{|\Gamma|} \sum_{p \in \Gamma} \frac{|H_1(p) - H_2(p)|}{H_{\max}},
\end{equation}
where $H_{\max}$ is a normalization constant representing the maximum expected curvature difference.
For volumetric alignment, we quantify volumetric overlap via a volumetric IoU score:
\begin{equation}
S_{\text{vol}}(C_1, C_2) = \frac{\text{Vol}(C_1) + \text{Vol}(C_2)}{\text{Vol}(\text{CH}(C_1 \cup C_2))},
\end{equation}
where $\text{CH}(C_1 \cup C_2)$ denotes the convex hull of the merged region. A higher $S_{\text{vol}}$ indicates better volumetric consistency between the parts.
The merging score is computed as a weighted combination:
\begin{equation}
S(C_1, C_2) = \beta \cdot S_{\text{curv}}(C_1, C_2) + \gamma \cdot S_{\text{vol}}(C_1, C_2),
\end{equation}
where $\beta$ and $\gamma$ control the relative importance of curvature and volume. Two parts are merged if $S(C_1, C_2) > \tau_m$, $\tau_m$ is a threshold.

\subsubsection{Block-Regrow-Fill}
\label{subsubsec:block_regrow_fill}

Our structurally guided shape abstraction starts with a key observation that each volumetric partition is an occupancy grid -- the voxels with one label as interior and all others as exterior. This yields the 3D shape for that label, and we can compute its corresponding SDF field.

\begin{figure}[H]
    \vspace{-0.2cm}
    \centering
    \includegraphics[width=\columnwidth]{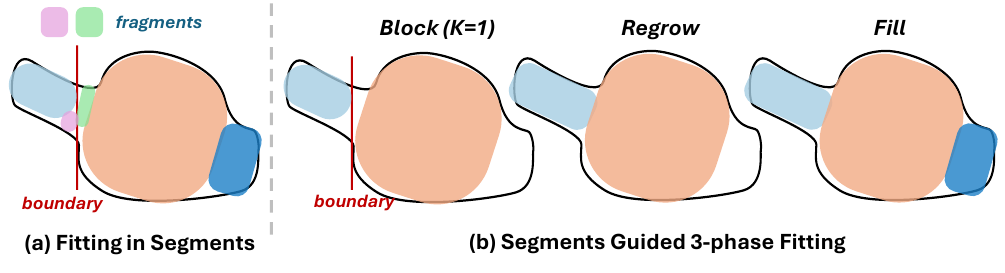}
    \vspace{-16pt}
    \caption{A 2D illustration of direct fitting in segments and ``block-regrow-fill''. The gaps around the boundary are filled through the regrow-stage.}
    \vspace{-0.4cm}
    \label{fig:three_stage_fitting}
\end{figure}

After extracting the shape for each region, a straightforward strategy is to fit superquadrics independently to each partition. Although this achieves structure-aware alignment, the axis-aligned cutting planes may form unnatural boundaries for superquadrics. More specifically, the algorithm will generate cluttered and meaningless superquadrics fragments along these boundaries to fill the shape, as shown in \cref{fig:three_stage_fitting}-(a), undermining the goal of shape abstraction. Therefore, we proposed a three-stage fitting strategy to provide greater flexibility at structural boundaries:

\noindent\textbf{1. Block.} Fit at most $K$ superquadrics to each segment.

\noindent\textbf{2. Regrow.} Insert all fitted superquadrics $\bm{\theta}_1,\cdots,\bm{\theta}_N$ into the original shape. For each superquadric $\bm{\theta}_i$, use it as the initialization and run a second optimization, to obtain the final state $\hat{\bm{\theta}}_i$. The target SDF field is computed from a series of SDF carving: 
\begin{equation}
    \phi_{\text{target}} = \phi \setminus \phi_{\hat{\bm{\theta}}_1} \setminus \cdots \setminus \phi_{\hat{\bm{\theta}}_{i-1}} \setminus \phi_{\bm{\theta}_{i+1}} \setminus \cdots \setminus \phi_{\bm{\theta}_N}
\end{equation}
Ideally, superquadrics in other regions can form flexible boundaries and allow ``regrow'' to fill the gaps around them.

\noindent\textbf{3. Fill.} Fill under-fitted regions with additional superquadrics.

The effectiveness of this algorithm depends on the proper choice of $K$, and whether the placeholder superquadrics can reliably guard the boundaries. Since the structural partitions from \cref{subsubsec:decomposition} exhibit strong convexity, we found that $K=1$ already yields ideal results.

\begin{figure}[t]
    \centering
    \includegraphics[width=\columnwidth]{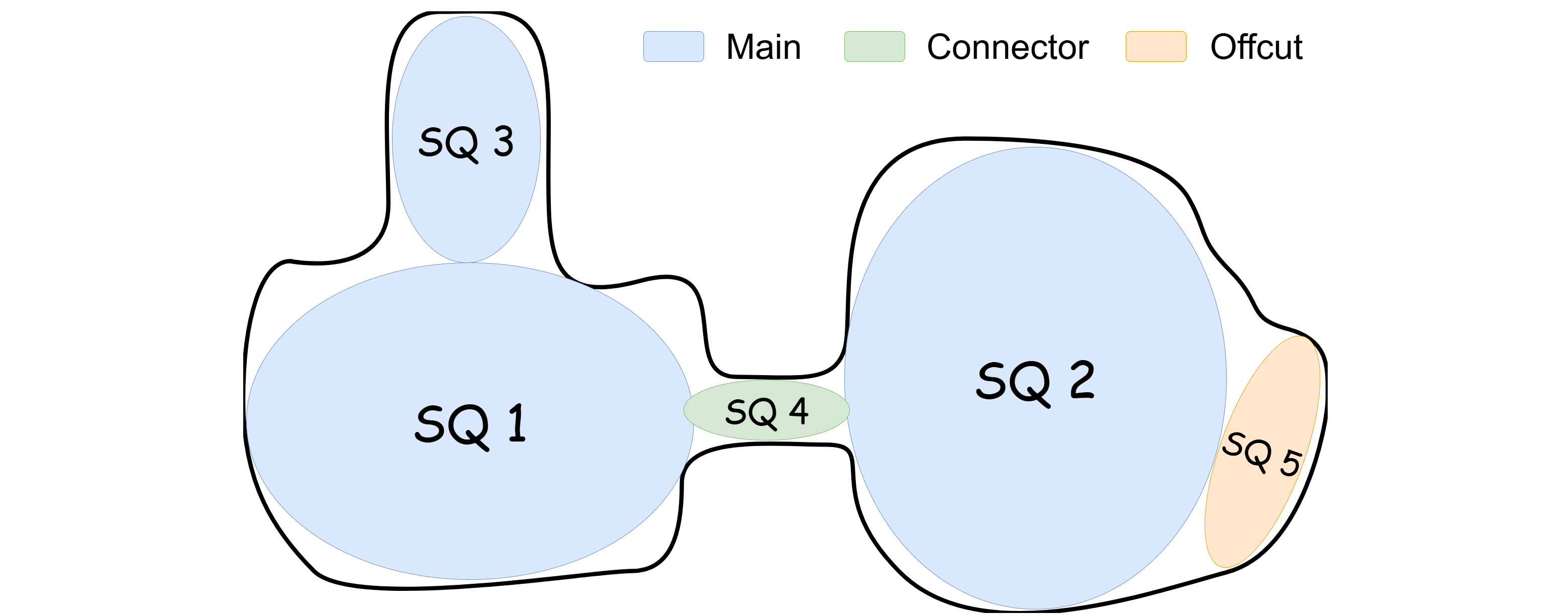}
    \caption{A 2D illustration of superquadric classification used in adaptive residual pruning. SQ 1–2 are fitted during the 'block' and 'regrow' stages, while SQ 3–5 are fitted during the 'fill' stage.}
    \vspace{-0.3cm}
    \label{fig:eps}
\end{figure}

\subsection{Adaptive Residual Pruning}
\label{subsec:pruning}

To enable more effective editing of primitive-based models after generation, a compact representation is essential. For instance, in a humanoid model, editing the "head" becomes much easier if it is represented by a single superquadric rather than multiple. 

Achieving such compactness hinges on deciding when to stop fitting superquadrics on the TSDF. In other words, we must determine whether to fit a new superquadric on each connected component. This is particularly important for generated meshes, which often lack the clean structural organization of artist-designed models. To preserve key geometric features while minimizing redundancy, small but meaningful components (e.g., the propeller of a plane, as shown in Fig. \ref{fig:teaser}) should be retained, while large but structurally insignificant artifacts (e.g., residuals from fitting a superquadric to a slightly tilted sphere) should be discarded.

The challenge then turns to how we can classify these components. An intuitive idea is that the superquadrics fitted during the ``block'' and ``regrow'' stages should be retained, since they each correspond to a distinct and meaningful structural partition. Therefore, we primarily focus on the connected components captured during the fill stage. For each remaining TSDF component, we initialize the superquadric as the maximal inscribed sphere, then classify the connected component into one of the following categories (see Fig. \ref{fig:eps} for a 2D illustration):

\noindent\textbf{Main SQ.} The majority of SDF values (over $P_M$\%) covered by the superquadric remain untouched, indicating an unfitted region.

\noindent\textbf{Connector SQ.} Most SDF values (over $P_C$\%) covered by this init superquadric are updated, and the updates are from more than one \emph{Main SQ}s. It represents a bridging part between \emph{Main SQ}s.

\noindent\textbf{Offcut SQ.} Most SDF values (over $P_O$\%) covered by this initial superquadric are updated by one and only one \emph{Main SQ},  corresponding to leftover regions.

We then apply ascending pruning thresholds $T_M$, $T_C$, and $T_O$ to the Main, Connector, and Offcut categories, respectively. If the minimum scale $min(a_x,a_y,a_z)$ of the initial superquadric is below the corresponding threshold, the superquadric is discarded, and the TSDF component is skipped. Detailed setting in \emph{supplementary} \cref{supp_sec:implementation}.

\subsection{Multiscale Fitting}
\label{subsec:multiscale}

\begin{figure}[t]
    \centering
    \includegraphics[width=\columnwidth]{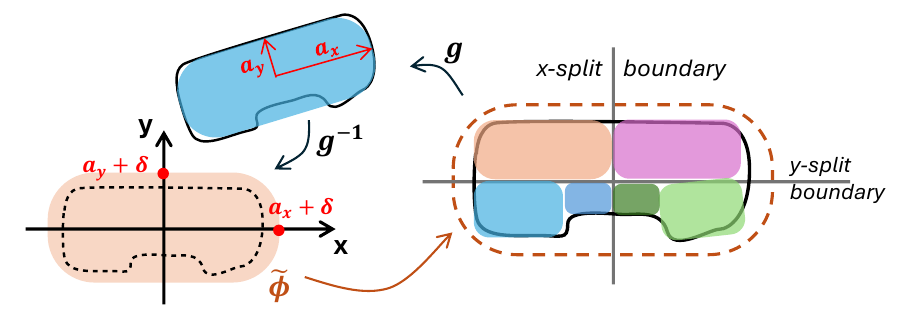}
    \vspace{-0.4cm}
    \caption{A 2D illustration of multiscale fitting. The region originally covered by a single superquadric is irregular. By splitting the retrieved $\tilde{\phi}$ for one time along the x-axis and y-axis, we obtain four partitions and fit a total of six superquadrics, which better capture the shape.}
    \vspace{-0.4cm}
    \label{fig:multiscale}
\end{figure}

Sometimes superquadric fitting over‐abstracts a local region, and when we seek to improve the surface fitting precision there, existing methods have largely been ineffective. In contrast, our method can upsample the region represented by a given superquadric. Assume the superquadric is parameterized by $\bm{\theta}=[\epsilon_1,\epsilon_2,a_x,a_y,a_z,g(\bm{e},\bm{t})]$, we first obtain its ``axis-aligned dilated version'' as
\begin{equation}
    \tilde{\bm{\theta}}=[\epsilon_1,\epsilon_2,a_x+\delta,a_y+\delta,a_z+\delta,g(\bm{0},\bm{0})]
    \label{eq:aadil}
\end{equation}
Here $\delta$ ensures that nearby pruned regions can also be covered by $\tilde{\bm{\theta}}$. Next, we can use grid sampling to capture the actual shape of the region represented by this superquadric. Specifically, for any point $\bm{x}$ within the axis-aligned $\tilde{\bm{\theta}}$, we apply the transformation $g$ to map it back to the original shape and sample the associated SDF value:
\begin{equation}
    \tilde{\phi}(\bm{x})\leftarrow \phi(g(\bm{x}))
\end{equation}
The orientation of the shape in $\tilde{\phi}$ is now better aligned with the coordinate axes. We then subdivide the field evenly along each axis to form volumetric partitions and fit superquadrics to each one. This partitioning provides higher‐resolution sampling of the space, enabling the capture of finer local structure, which is further illustrated in \cref{fig:multiscale}.
\section{3DGen-Prim Dataset}
\label{sec:3dgenprim}

Since our algorithm targets an image-to-primitives pipeline in UGC scenarios, we evaluate it on image-to-3D generation outputs. We use the 510 image prompts provided by 3DGen-Bench~\cite{zhang20253dgenbench} and generate test meshes with two state-of-the-art image-to-3D methods, Hunyuan3D-2.0~\cite{zhao2025hunyuan3d2} and TripoSG~\cite{li2025triposg}. The outputs of both methods are extracted from SDF or occupancy fields, yielding excellent watertightness, well-suited for testing our approach.

In our evaluation, we first measure how well the abstraction fits the input shape using three metrics: Chamfer Distance (CD), Earth Mover’s Distance (EMD), and Voxel-IoU. CD and EMD are computed on surface point clouds. Since primitive decomposition can introduce self-occlusion between parts, we sample points via a rasterized scanning procedure. For Voxel-IoU, we extract the “inside voxels” of the input and output and compute their intersection over union. Notably, PrimitiveAnything~\cite{ye2025primitiveanything} also reports Voxel-IoU, but only extracts shape surface voxels, which admits ambiguity. In contrast, our method can distinguish ``fake shape abstraction'' which merely lines the surface with thin primitives.

We also evaluate \emph{editability} against the UGC standards outlined in \cref{sec:intro}. For low overlap, we compute the Overlap Rate (OR) based on the extracted inside voxels. Given a primitive $\bm{\theta}$, $M_{\bm{\theta}}$ is the mask that indicates whether each voxel is inside this primitive. We compute the average number of primitives that each voxel is covered by as:
\begin{equation}
    \mathrm{OR}=\frac{\sum_{\bm{x}} \left( \sum_{\bm{\theta}}M_{\bm{\theta}} \right)}{\sum_{\bm{x}} \left( \bigcup_{\bm{\theta}}M_{\bm{\theta}} \right)}
\end{equation}
For compactness, when the fitting accuracy is comparable, the average number of primitives $\bar{N}$ serves as an indicator of how well surface fragments are cleaned up.
\section{Experiments}
\label{sec:exp}

\subsection{Generated Shape Abstraction}
\label{subsec:shape_abstraction}

\begingroup
\setlength{\tabcolsep}{5.6pt}
\begin{table*}[t]
\caption{\textbf{Quantitative Comparison.} We highlight the best value in \colorbox{pearDark!20}{blue}, and the second-best value in \colorbox{mycolor_green}{green}. ``Optim.'', ``Learn.'', and ``Rule'' stand for optimization-based, learning-based, and rule-based approaches, respectively. We do not highlight the best $\bar{N}$ because its comparison only makes sense when two methods achieve similar fitting performance; otherwise, a small $\bar{N}$ simply indicates underfitting to the input geometry.}
\vspace{-8pt}
{\fontsize{8.2pt}{11pt}\selectfont
\begin{tabular}{ll|ccc|c|c|ccc|c|c}
\toprule
\multirow{2}{*}{Method}   & \multirow{2}{*}{Type} & \multicolumn{5}{c|}{\textbf{Hunyuan3D-2.0}}  & \multicolumn{5}{c}{\textbf{TripoSG}} \\ 
                          &                        &  CD $\downarrow$ & EMD $\downarrow$ & Voxel-IoU $\uparrow$ & OR $\downarrow$ & $\bar{N}$ & CD $\downarrow$ & EMD $\downarrow$ & Voxel-IoU $\uparrow$ & OR $\downarrow$ & $\bar{N}$                   \\ \midrule

EMS \cite{liu2022emsrobust} & Optim. & 0.2345 & 0.2036 & 0.466 & 1.524 & 3.44 & 0.2472 & 0.2168 & 0.436 & 1.540 & 3.63 \\
Marching-Primitives \cite{liu2023marching}  & Optim. & \tablesecond{0.0396} & \tablesecond{0.0544} & \tablefirst{0.868} & 4.201 & 67.7 & \tablesecond{0.0403} & \tablefirst{0.0537} & \tablesecond{0.860} & 3.778 & 62.5 \\
AISSR \cite{li2025aissr} & Rule & 0.1128 & 0.0918 & 0.403 & \tablesecond{1.051} & 7.94 & 0.1132 & 0.0915 & 0.394 & \tablesecond{1.056} & 7.81 \\
PrimitiveAnything \cite{ye2025primitiveanything} & Learn. & 0.1366 & 0.0986 & 0.442 & 1.845 & 82.7 & 0.1299 & 0.0936 & 0.457 & 1.870 & 79.9 \\ \midrule
Light-SQ (Ours) & Optim. & \tablefirst{0.0388} & \tablefirst{0.0531} & \tablesecond{0.861} & \tablefirst{1.015} & 61.0 & \tablefirst{0.0385} & \tablesecond{0.0538} & \tablefirst{0.864} & \tablefirst{1.016} & 62.1 \\ \bottomrule
\end{tabular}
}
\label{tab:quant_main}
\vspace{-0.2cm}
\end{table*}
\endgroup

\begin{figure*}[t]
    \centering
    \includegraphics[width=0.95\textwidth]{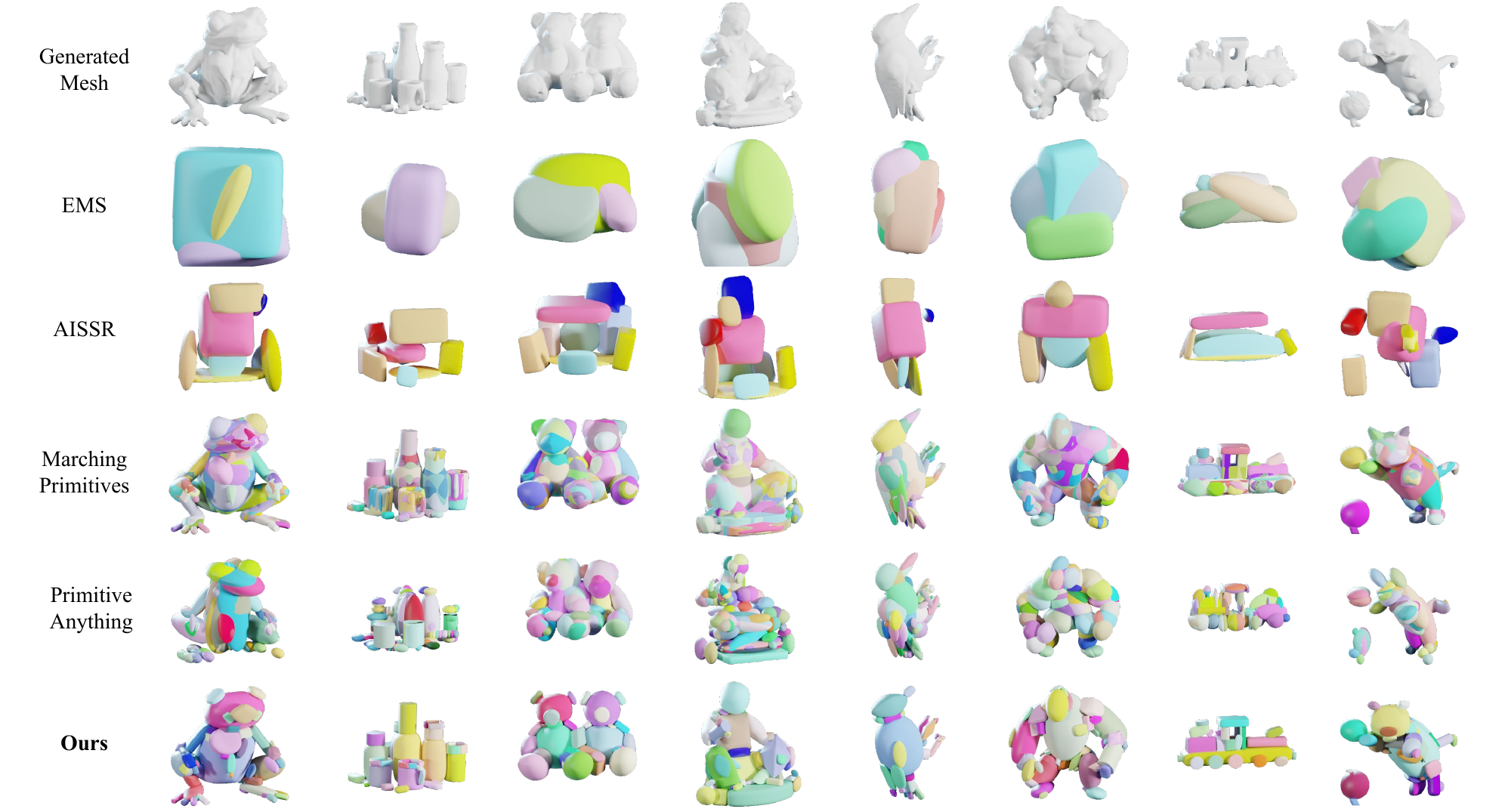}
    \vspace{-8pt}
    \caption{\textbf{Qualitative Comparison.} Our method is the only one that simultaneously achieves faithful fitting to the original shape and strong editability.}
    \label{fig:main_qualitative}
\end{figure*}

\noindent\textbf{Comparison Methods.} We compare our approach to EMS~\cite{liu2022emsrobust} and Marching-Primitives~\cite{liu2023marching} among optimization-based methods, which both use superquadrics as the primitive representation. Since Marching-Primitives also uses TSDF as its fitting target, we set the input SDF field resolution for both methods to $100^3$ for a fair comparison. Furthermore, among rule-based and learning-based methods, we compare against the current state-of-the-art, AISSR~\cite{li2025aissr} and PrimitiveAnything~\cite{ye2025primitiveanything}, respectively. AISSR produces three versions of shape abstraction per run. We select the instance‐level output (AISSR-I), which does not impose semantic‐level constraints on the fitting of deformable superquadrics, thereby granting maximal freedom and yielding the best performance in metrics and visual quality.

\noindent\textbf{Quantitative Results.} We show the quantitative comparison on our 3DGen-Prim dataset in \cref{tab:quant_main}. EMS significantly underperforms on fitting metrics because, at its core, it is still an algorithm for fitting a single primary superquadric and is not well suited to handling complex geometric structures. Marching-Primitives achieves state‐of‐the‐art performance across all fitting metrics, as it always strives to completely fill a shape’s interior. However, its fitted primitives overlap excessively that, on both datasets, each voxel is covered by an average of four superquadrics, predictably making the results difficult to edit. AISSR produces highly editable results, but it consistently fits only a small number of primitives, limiting its ability to represent complex shapes, which is a common shortcoming of rule-based methods. PrimitiveAnything, despite being trained on a 120K-sample dataset, still shows clear generalization issues on our generative data, offering no advantage in fitting quality or editability. Our method achieves fitting quality on par with Marching-Primitives, significantly outperforms other approaches, and simultaneously attains the lowest overlap—ensuring superior editability.

\noindent\textbf{Qualitative Results.} \cref{fig:main_qualitative} presents the qualitative comparison. The generated mesh is randomly sampled from our 3DGen-Prim dataset. EMS and AISSR exhibit significant generalization challenges in fitting quality. Marching-Primitives achieves high reconstruction fidelity to the input shape, but the primitives overlap excessively and fail to reflect the underlying structure, resulting in poor editability. As a learning‐based method, PrimitiveAnything demonstrates remarkable generalization, but its fitting quality is unstable, and on more complex generated shapes, it tends to do crude stacking, which harms editability. Our method achieves stable, high-precision fitting to the input geometry, while effectively ensuring low overlap and structure-aware alignment for editability.

\begingroup
\setlength{\tabcolsep}{10.8pt}
\begin{table}[t]
\caption{\textbf{Editability User Study.} We show the average ranking of each method in every metric. ``PrimAny'' stands for PrimitiveAnything \cite{ye2025primitiveanything}. ``MPS'' stands for Marching-Primitives \cite{liu2023marching}.}
\vspace{-8pt}
{\fontsize{8.2pt}{11pt}\selectfont
\begin{tabular}{l|ccc}
\toprule
Metric & PrimAny & MPS & Ours \\ \midrule
Geometry Editability & 2.43 & 2.50 & \textbf{1.07} \\
Geometry Editing Efficiency & 2.46 & 2.51 & \textbf{1.03} \\ \midrule
Texture Editability & 2.32 & 2.61 & \textbf{1.06} \\
Texture Editing Efficiency & 2.32 & 2.62 & \textbf{1.06} \\ \midrule
Animation Friendliness & 2.34 & 2.61 & \textbf{1.04} \\ 
\bottomrule
\end{tabular}
}
\label{tab:main_user_study}
\vspace{-0.2cm}
\end{table}
\endgroup

\noindent\textbf{User Study on Editability.} We conducted an extended user study for two baselines, Marching-Primitives and PrimitiveAnything, and our method on five metrics: Geometry Editability, Geometry Editing Efficiency, Texture Editability, Texture Editing Efficiency, and Animation Friendliness. 
In our user study, 25 participants received eight sets of uncurated shape abstraction results and were instructed to rank each set according to the five aforementioned criteria. The backgrounds of the participants included game development engineers and asset creation artists.
They were provided with clear explanations and examples of each metric before evaluation, ensuring a consistent interpretation across raters. The explanations are detailed in \emph{supplementary} \cref{supp_sec:user_study}. Their ratings showed good agreement, suggesting that the metrics are interpretable and meaningful from a user’s perspective.
\cref{tab:main_user_study} reports the average ranking of different methods on each metric. Our method exhibits a comprehensive, across‑the‑board advantage.

\begin{figure*}[t]
    \centering
    \includegraphics[width=0.6\textwidth]{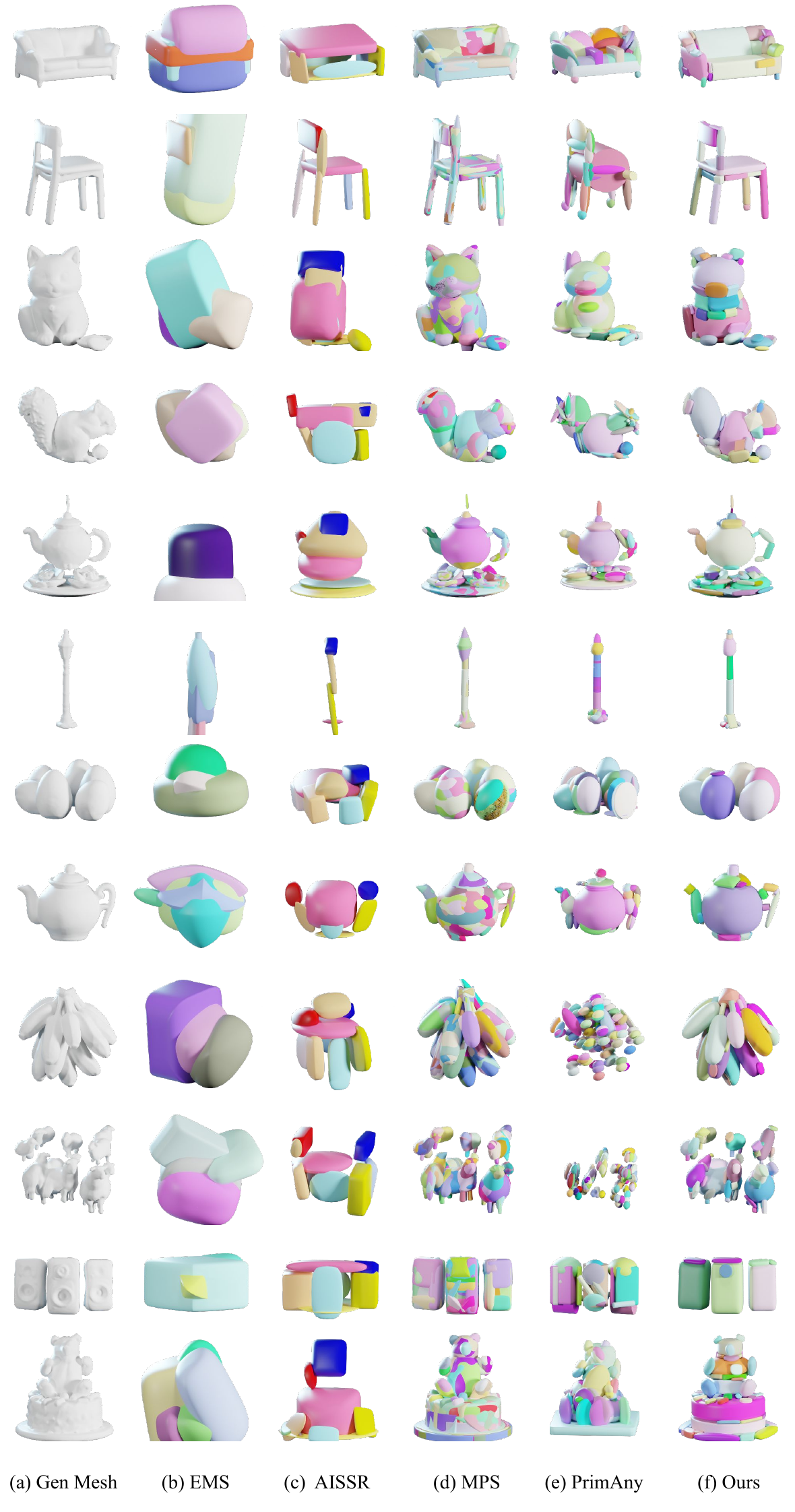}
    \caption{Additional results on our 3DGen-Prim dataset, comparing (b) EMS~\cite{liu2022emsrobust}, (c) AISSR~\cite{li2025aissr}, (d) Marching-Primitives~\cite{liu2023marching}, (e) PrimitiveAnything~\cite{ye2025primitiveanything}, and (f) Our Light-SQ.}
    \label{fig:more_results}
\end{figure*}

\noindent\textbf{Efficiency Comparison.} We benchmarked the speed of all methods on a single-GPU (14592 CUDA cores, 96GB VRAM) workstation with AMD EPYC 9K84 96‑core CPU. Our method completes in 25.98 seconds per shape, including preprocessing, which is over 10× faster than Marching-Primitives (339.59s) and even slightly faster than PrimitiveAnything (29.10s), when accounting for its point-cloud scanning overhead. AISSR is very fast (0.06s), but its fitting fidelity is clearly insufficient (see \cref{fig:main_qualitative} and \cref{fig:more_results}). EMS provides no advantage in either speed (19.05s) or fitting quality.

\subsection{Ablation Study and Discussions}
\label{subsec:ablation}

\begingroup
\setlength{\tabcolsep}{4.3pt}
\begin{table}[t]
\caption{\textbf{Ablation on Low-overlapping.} Results are shown in \texttt{OR (N)}. Here \texttt{OR} and \texttt{N} stand for overlap rate and average number of primitives, respectively. $\tau$ is the TSDF truncation.}
\vspace{-8pt}
{\fontsize{8.2pt}{11pt}\selectfont
\begin{tabular}{l|cccc}
\toprule
 & $w=0.50$ & $w=0.10$ & $\bm{w=0.02}$ & $w=0.01$ \\ \midrule
$C=0.1$ & 1.019 (45.2) & 1.018 (50.1) & 1.016 (51.4) & 1.016 (55.8) \\
$\bm{C=1}$ & 1.025 (51.1) & 1.016 (59.5) & \textbf{1.014 (61.7)} & 1.010 (79.4) \\
$C=1/\tau=50$ & 1.012 (75.2) & 1.011 (91.0) & 1.012 (113.8) & 1.012 (117.8) \\
\bottomrule
\end{tabular}
}
\label{tab:abl_or_n}
\vspace{-0.2cm}
\end{table}
\endgroup
\begingroup
\setlength{\tabcolsep}{7pt}
\begin{table}[t]
\caption{User study results comparing our method with the ablated version and PrimitiveAnything, in terms of structure-aware alignment.}
\vspace{-8pt}
{\fontsize{8.2pt}{11pt}\selectfont
\begin{tabular}{l|ccc|c}
\toprule
Method & Rank-1 & Rank-2 & Rank-3 & Avg. Rank \\ \midrule
PrimitiveAnything & 6\% & 12\% & 82\% & 2.76 \\
w/o. structure-aware & 15\% & 74\% & 10\% & 1.96 \\
Ours & 75\% & 16\% & 9\% & 1.35 \\ \bottomrule
\end{tabular}
}
\label{tab:user_study}
\vspace{-0.2cm}
\end{table}
\endgroup
\begingroup
\setlength{\tabcolsep}{7.2pt}
\begin{table}[t]
\caption{\textbf{Block-Regrow-Fill Ablation.} Ablating the necessity of our 3-phase fitting strategy. ``CH'' stands for convex hull.}
\vspace{-8pt}
{\fontsize{8.2pt}{11pt}\selectfont
\begin{tabular}{l|ccc|c|c}
\toprule
Method   &  CD $\downarrow$ & EMD $\downarrow$ & Voxel-IoU $\uparrow$ & OR $\downarrow$ & $\bar{N}$ \\ \midrule
One-per-CH    & 0.0613 & 0.0634 & 0.735 & 1.018 & 20.9 \\
Ours          & \textbf{0.0388} & \textbf{0.0531} & \textbf{0.861} & 1.015 & 61.0 \\ \bottomrule
\end{tabular}
}
\label{tab:abl_one_per_ch}
\vspace{-0.2cm}
\end{table}
\endgroup
\begingroup
\setlength{\tabcolsep}{5.4pt}
\begin{table}[t]
\caption{\textbf{Residual Pruning Ablation.} Pruning has almost no impact on fitting accuracy, yet it significantly reduces the number of primitives.}
\vspace{-8pt}
{\fontsize{8.2pt}{11pt}\selectfont
\begin{tabular}{l|ccc|c|c}
\toprule
\textbf{Hunyuan3D-2.0}   &  CD $\downarrow$ & EMD $\downarrow$ & Voxel-IoU $\uparrow$ & OR $\downarrow$ & $\bar{N}$ \\ \midrule
w/o. pruning    & 0.0367 & 0.0533 & 0.874 & 1.018 & \textbf{95.4} \\
Ours                     & 0.0388 & 0.0531 & 0.861 & 1.015 & \textbf{61.0} \\ \bottomrule
\textbf{TripoSG}   &  CD $\downarrow$ & EMD $\downarrow$ & Voxel-IoU $\uparrow$ & OR $\downarrow$ & $\bar{N}$ \\ \midrule
w/o. pruning    & 0.0371 & 0.0526 & 0.872 & 1.018 & \textbf{92.2} \\
Ours            & 0.0385 & 0.0538 & 0.864 & 1.016 & \textbf{62.1} \\ \bottomrule
\end{tabular}
}
\label{tab:abl_residual}
\vspace{-0.2cm}
\end{table}
\endgroup

\noindent\textbf{Low overlapping.} While our method does not guarantee zero overlap, it significantly reduces it via SDF carving and the penalty. Here, we provide an ablation on $w$ and $C$ in \cref{eq:lambda}. We conducted a grid test for these two parameters on 51 uncurated Hunyuan3D-2.0 \cite{zhao2025hunyuan3d2} samples in our 3DGen-Prim dataset, comparing the overlap rate and the average number of primitives. \cref{tab:abl_or_n} shows that lower $w$ and higher $C$ reduce overlap rate (OR) but increase fragmentation. The paper setting is highlighted.

\begin{figure*}[htb]
\centering
\begin{minipage}[t]{0.45\textwidth}
    \centering
    \includegraphics[width=1\linewidth]{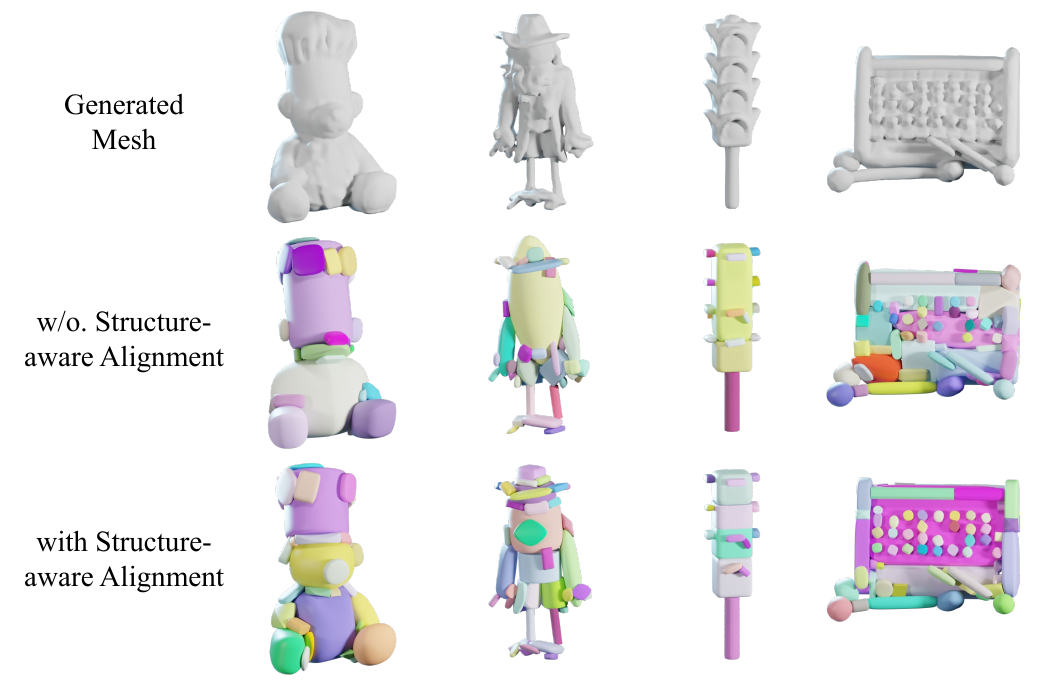}
    \caption{\textbf{Ablation on Structure-aware Alignment.} Without structure-aware alignment, superquadrics fitted in early stages may span multiple structural partitions, resulting in degraded visual quality and editability.}
    \label{fig:structure_ablation}
\end{minipage}
\quad
\begin{minipage}[t]{0.52\textwidth}
    \centering
    \includegraphics[width=1\linewidth]{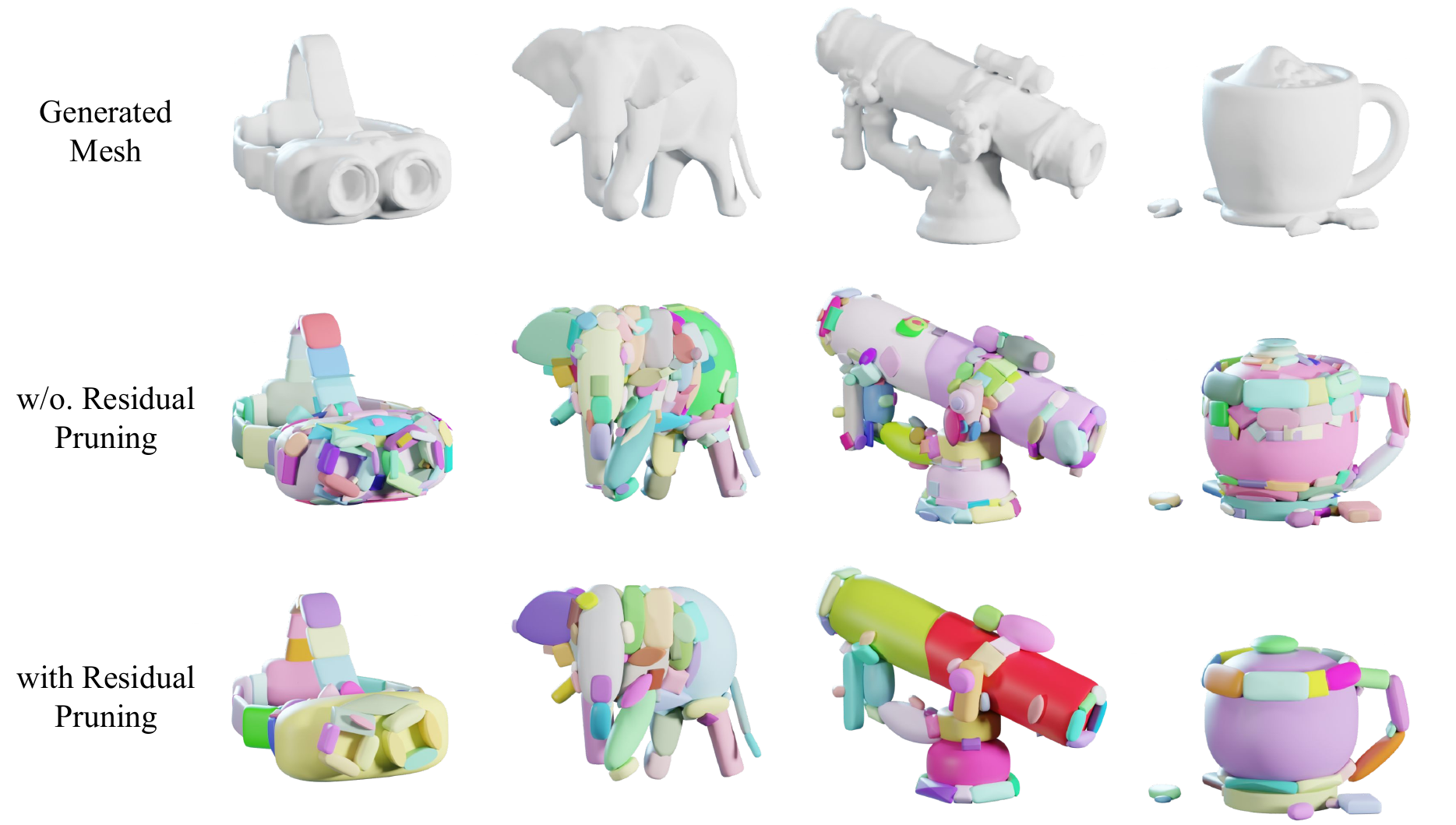}
    \caption{\textbf{Ablation on Residual Pruning.} Since optimization-based fitting algorithms always try to fully occupy the interior of the input geometry, they tend to generate many fragmented superquadrics around the surface. Adaptive residual pruning effectively removes these fragments.}
    \label{fig:residual_ablation}
\end{minipage}
\end{figure*}

\noindent\textbf{Structure-aware Alignment.} We compare our method with an ablated version without structure-aware convex decomposition and ``block-regrow-fill'' strategy. \cref{fig:structure_ablation} demonstrates how these two modules help our algorithm perceive and align with the input structure. Since this objective is difficult to quantify, we conducted a user study in which 28 participants were asked to rank 11 sets of randomly selected shape abstraction results. Results from PrimitiveAnything are also included. The participants were instructed to pick the best and worst primitive‑decomposition results based on whether the structural partitioning was clear and reasonable. The comparisons are anonymized, and the order is randomized. The results shown in \cref{tab:user_study} further demonstrate the significant impact of our convex decomposition method and fitting strategy on enhancing structure-aware performance.

\noindent\textbf{Block-Regrow-Fill.} As mentioned in \cref{subsubsec:block_regrow_fill}, given the structural guidance of a set of pseudo convex hulls, a straightforward approach is to fit superquadrics independently to each partition. To show the necessity of our three-phase fitting strategy, we compare our results with fitting a single superquadric to each pseudoconvex hull. As is showin in \cref{tab:abl_one_per_ch}, since not all convex components can be well represented by a single superquadric, the ablated variant exhibits a pronounced drop in the fitting metrics.

\noindent\textbf{Residual Pruning.} \cref{fig:residual_ablation} ablates the effect of adaptive residual pruning. The quantitative comparison in \cref{tab:abl_residual} further shows that, while the other metrics remain identical, pruning reduces the primitive count by an average of 30, thus reducing surface fragmentation.

\begin{figure*}[t]
    \centering
    \includegraphics[width=\textwidth]{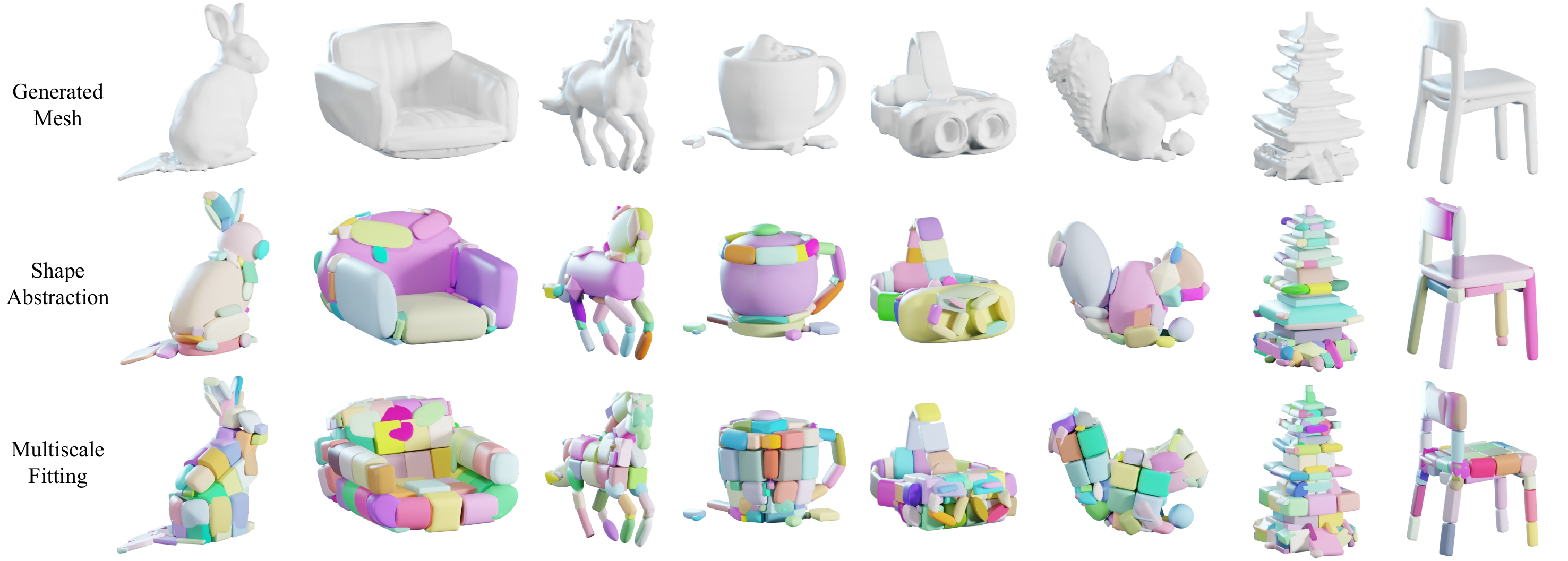}
    \caption{\textbf{Multiscale Fitting.} Based on the results of shape abstraction, multiscale fitting can upsample certain spatial regions and capture finer local curvature features in a block‐stacking fashion. Users can refine the fitting of a specific region by selecting the corresponding superquadric.}
    \label{fig:multiscale_qualitative}
\end{figure*}

\noindent\textbf{Multiscale Fitting.} Qualitative results are shown in \cref{fig:multiscale_qualitative}. Multiscale fitting supports spatial upsampling and shape refinement, enabling the capture of finer local structure.
\section{Conclusion}
\label{sec:conclusion}

We introduce Light-SQ, a structure-aware superquadric decomposition framework for generated 3D geometry in UGC scenarios. Using SDF carving, structure-aware alignment, and adaptive residual pruning, Light-SQ yields abstractions that are simultaneously compact, editable, and faithful to the original geometry. We perform extensive experiments using our constructed 3DGen-Prim dataset to benchmark against prior methods, demonstrating consistent advantages in both fidelity and structural usability.

\vspace{0.21cm}

\small\noindent\textbf{Acknowledgement.} This work is supported by the National Research Foundation, Singapore under its AI Singapore Programme (AISG Award No: AISG2-PhD-2022-01-030).

\clearpage

\clearpage

\clearpage
\bibliographystyle{ACM-Reference-Format}
\bibliography{light_sq_ref}

\clearpage
\appendix
\section{Implementation Details}
\label{supp_sec:implementation}

In this section, we summarize the implementation details of our structure-aware superquadrics fitting algorithm.

\noindent\textbf{Non-overlapping Fitting.} The TSDF matching term follows a normal distribution as:
\begin{equation}
    P \left( \phi_{\tau}(\bm{x})|\phi_{\bm{\theta},\tau}(\bm{x}) \right) = \mathcal{N} \left( \phi_{\tau}(\bm{x}) | \phi_{\bm{\theta},\tau}(\bm{x}), \sigma^2 \right)
    \label{eq:tsdf_matching}
\end{equation}
The $\sigma^2$ parameter is updated in closed form through the EM optimization process of the superquadric fitting. We further apply a truncation for numerical stability:
\begin{equation}
    \sigma^2= \max \left( \frac{\sum_{\bm{x}\in\bm{V}_a(\bm{\theta})} \lambda_{\bm{x}}\|\phi_{\bm{\hat{\theta}},\tau}(\bm{x})-\phi_{\tau}(\bm{x})\|_2^2}{\sum_{\bm{x}\in\bm{V}_a(\bm{\theta})} \lambda_{\bm{x}}}, \tau^2\right)
\end{equation}
For the $\lambda_{\bm{x}}$ computation scheme, we set the primitive number prior $\tilde{N}=50$, which yields $w=1/\tilde{N}=0.02$. The weighting constant $C$ is set to $1$. Compared to previous methods~\cite{liu2023marching}, our inside-voxel decay term is much smaller (yet still much larger than the TSDF matching term), allowing more flexible fitting around the generated shape surface, since the generated surface is not as clean as the artist-made ones.

\noindent\textbf{Structure-aware Convex Decomposition.} For the slice-plane generation, we operate on the axis-aligned setting, thus given a $100^3$ grid, there are 300 potential cutting planes in total. We set $\alpha=0.7$ for the structural saliency score, $K=6$ and $\delta=0.1$ to finalize the candidates. This setting also generalizes to other grid resolutions, given that the input geometry is normalized to $[-1,1]^3$.

For adaptive merging, we set $\beta=0.4$ and $\gamma=0.6$ to balance between curvature and volumetric score. The merging threshold $\tau_m$ is set to $0.7$.

\noindent\textbf{Block-Regrow-Fill.} During ``block'' stage, we set $K=1$. The optimization parameters remain the same through all three stages as the setting above.

\noindent\textbf{Adaptive Residual Pruning.} In practice, we find that simply setting the classification percentages $P_M\%=P_C\%=P_O\%=50\%$ yields robust classification. For the pruning threshold, we normalize all generated geometry into $[-1,1]^3$ and set thresholds $T_M=V=0.02$, $T_C=1.5V=0.03$, and $T_O=2.5V=0.05$, where $V=2/100=0.02$ is the TSDF voxel size.

\section{User Study Settings}
\label{supp_sec:user_study}

Here we detail the explanation of the five metrics in main paper Table 2, which are also provided to the participants.

\begin{enumerate}
    \item \emph{Geometry Editability.} Whether the structural components of the model are clearly distinguishable and can be locally modified without unintended influence on other parts.
    \item \emph{Geometry Editing Efficiency.} Whether the structure facilitates rapid editing and ensures that geometric modifications accurately reflect the intended target shape.
    \item \emph{Texture Editability.} Whether texture regions are clearly defined and can be modified independently without affecting unrelated areas.
    \item \emph{Texture Editing Efficiency.} Whether target texture regions can be quickly identified and edited in a manner consistent with the user’s intention and alignment.
    \item \emph{Animation Friendliness.} Whether the model can be easily rigged with a skeleton while maintaining natural and plausible deformations during local transformations.
\end{enumerate}

\end{document}